\documentclass[12pt]{article}

\catcode`@=12
\begin{document}
\thispagestyle{empty}
\begin{flushright} 
UCRHEP-T336\\ 
May 2002\
\end{flushright}
\vspace{0.5in}
\begin{center}
{\LARGE	\bf Supersymmetric Higgs Triplets and\\ Bilinear R-Parity 
Nonconservation\\}
\vspace{1.5in}
{\bf Ernest Ma\\}
\vspace{0.2in}
{\sl Physics Department, University of California, Riverside, 
California 92521\\}
\vspace{1.5in}
\end{center}
\begin{abstract}\
The supersymmetric standard model of particle interactions is extended to 
include two Higgs triplet superfields at the TeV scale, carrying two units 
of lepton number.  Realistic tree-level Majorana neutrino masses are obtained 
in the presence of soft, i.e. bilinear, $R$-parity nonconservation.
\end{abstract}

\newpage
\baselineskip 24pt

In the minimal supersymmetric standard model (MSSM) of particle interactions, 
neutrinos are massless, because $R$-parity, i.e.~$R \equiv (-1)^{3B+L+2j}$, 
has been imposed.  If this is $softly$ broken in the superpotential, 
i.e.~using the bilinear terms $\mu_i \hat L_i \hat H_2$, then $one$ linear 
combination of the three doublet neutrinos gets a small effective seesaw 
Majorana mass from the $7 \times 7$ neutralino-neutrino mass matrix 
\cite{chfe}.  The other two linear combinations may also become massive, but  
only through radiative corrections \cite{loop}.  Another mechanism 
for obtaining small neutrino masses is to add a Higgs triplet $\xi = (\xi^{++},
\xi^+,\xi^0)$ with a small vacuum expectation value \cite{scva,masa}.  
However, since $\langle \xi^0 \rangle$ should be of order 1 eV or less, 
this is difficult to achieve with $m_\xi$ of order 1 TeV unless some new 
physics such as large extra dimensions is invoked \cite{marasa}.  On the 
other hand, in the presence of bilinear $R$-parity nonconservation, $\langle 
\xi^0 \rangle$ can be naturally small, as shown below.  The result is a 
simple supersymmetric model with three tree-level Majorana neutrino masses 
(without corresponding heavy singlet neutrino superfields) and no heavy 
particles beyond the TeV energy scale.

Consider the superfield particle content of the MSSM under the standard gauge 
group $SU(3)_C \times SU(2)_L \times U(1)_Y$, i.e.
\begin{eqnarray}
&& \hat Q_i = (\hat u_i, \hat d_i) \sim (3,2,1/6), ~~ \hat u_i^c \sim 
(3^*,1,-2/3), ~~ \hat d_i^c \sim (3^*,1,1/3), \\ 
&& \hat L_i = (\hat \nu_i, \hat e_i) \sim (1,2,-1/2), ~~ \hat e_i^c \sim 
(1,1,1), \\ 
&& \hat H_1 = (\hat \phi_1^0, \hat \phi_1^-) \sim (1,2,-1/2), ~~ \hat H_2 = 
(\hat \phi_2^+, \hat \phi_2^0) \sim (1,2,1/2).
\end{eqnarray}
The conservation of baryon number $B$ and lepton number $L$ is implicit in the 
above, which serves to distinguish $\hat L_i$ (with $L=1$) from $\hat H_1$ 
(with $L=0$).  Now add two superfields
\begin{equation}
\hat \xi_1 = (\hat \xi_1^{++}, \hat \xi_1^+, \hat \xi_1^0) \sim (1,3,1), 
~~ \hat \xi_2 = (\hat \xi_2^0, \hat \xi_2^-, \hat \xi_2^{--}) \sim (1,3,-1),
\end{equation}
with $L=-2$ and +2 respectively.  The superpotential of this model is then 
given by
\begin{eqnarray}
W &=& \mu \hat H_1 \hat H_2  + f^e_{ij} \hat H_1 \hat L_i \hat e_j^c + 
f^d_{ij} \hat H_1 \hat Q_i \hat d_j^c + f^u_{ij} \hat H_2 \hat Q_i \hat u_j^c 
\nonumber \\ &+& M \hat \xi_1 \hat \xi_2 + h_{ij} \hat L_i \hat L_j 
\hat \xi_1 + \mu_i \hat L_i \hat H_2,
\end{eqnarray}
where the $\mu_i$ terms break $L$ softly.  Note that all other terms which 
violate $L$, such as $\hat L_i \hat L_j \hat e_k^c$ and $\hat H_1 \hat H_1 
\hat \xi_1$, are trilinear and thus assumed to be absent in this model.

Because of the $\mu_i$ terms, the scalar neutrinos necessarily also acquire 
small vacuum expectation values \cite{drv}.  To see this, consider the 
following relevant terms in the Higgs potential:
\begin{equation}
V = \tilde m_L^2 \tilde \nu_i^* \tilde \nu_i + (\mu_i B \tilde \nu_i \phi_2^0 
+ H.c.) + ...,
\end{equation}
where the parameters $\tilde m_L$ and $B$ are the usual ones assumed in the 
soft breaking of the supersymmetry.  Let $u_i = \langle \tilde \nu_i \rangle$ 
and $v_2 = \langle \phi_2^0 \rangle$, then the minimization of $V$ yields 
\cite{ma01}
\begin{equation}
u_i \simeq -{\mu_i B v_2 \over \tilde m_L^2}.
\end{equation}
As a result, a linear combination of $\nu_i$ gets a Majorana mass given by 
\cite{hamasa}
\begin{equation}
m_\nu \simeq -\epsilon^2 \left( {s^2 \over M_1} + {c^2 \over M_2} \right),
\end{equation}
where $s = \sin \theta_W$, $c = \cos \theta_W$, and $M_{1,2}$ are the 
supersymmetry breaking Majorana masses of the $U(1)$ and $SU(2)$ gauginos 
respectively.  The parameter $\epsilon$ is given by
\begin{equation}
\epsilon_i = {M_Z \over v} \left( u_i - {\mu_i \over \mu} v_1 \right),
\end{equation}
with $\epsilon^2 = \epsilon_1^2 + \epsilon_2^2 + \epsilon_3^2$ and $v^2 = 
v_1^2 + v_2^2$.

Because of the $h_{ij}$ terms of Eq.~(5), the Lagrangian of this model has 
the following additional terms which are relevant for neutrino mass:
\begin{equation}
{\cal L} = (h_{ij} \nu_i \nu_j \xi_1^0 + h_{ij} A \tilde \nu_i \tilde \nu_j 
\xi_1^0 + H.c.) + M^2 {\xi_1^0}^* \xi_1^0 + ...
\end{equation}
Hence $\xi_1^0$ also gets a small vacuum expectation value given by
\begin{equation}
\langle \xi_1^0 \rangle \simeq - {h_{ij} A u_i u_j \over M^2},
\end{equation}
from which
\begin{equation}
({\cal M}_\nu)_{ij} = 2 h_{ij} \langle \xi_1^0 \rangle
\end{equation}
is obtained.  Note that the overall magnitude of this neutrino mass matrix is 
comparable to that of Eq.~(8) for $A$, $M$, and $M_{1,2}$ roughly of the 
same order of magnitude.  Combining the two allow for a general $3 \times 3$ 
neutrino mass matrix at tree level.

Without further assumptions, the texture of the neutrino mass matrix cannot 
be determined.  However, because $M$ is assumed to be of order 1 TeV, 
the production and decay of $\xi_1^{++}$ \cite{marasa} as well as other 
phenomenological predictions of bilinear $R$-parity violation \cite{drv} 
will be testable at future accelerators.

In conclusion, the supersymmetric standard model of particle interactions 
has been extended to include two Higgs superfields at the TeV scale, carrying 
two units of lepton number, together with the bilinear breaking of $R$-parity. 
This results in three naturally small tree-level Majorana neutrino masses 
with no new particles beyond the TeV scale, which renders this model 
experimentally verifiable.
\\[5pt]

This work was supported in part by the U.~S.~Department of Energy under 
Grant No.~DE-FG03-94ER40837.

\newpage
\bibliographystyle{unsrt}

\end{document}